\documentclass[aps,prl, 10pt, twocolumn,superscriptaddress]{revtex4-2}
\usepackage{multirow,eurosym,amssymb,amsfonts,setspace,graphicx,bm,float,amssymb}
\usepackage{xcolor,multirow,eurosym,amssymb,amsfonts,setspace,graphicx,bm,amssymb}
\usepackage{algorithmicx}
\usepackage{algpseudocode}
\usepackage[linesnumbered,ruled,vlined]{algorithm2e}
\usepackage{mathtools}

\usepackage{booktabs} 
\usepackage{color}

\usepackage{amsmath}

\newcommand{\bra}[1]{\langle #1|}
\newcommand{\ket}[1]{|#1\rangle}

\newcommand{\sol}[1]{} 

\def\be{\begin{equation}}
\def\ee{\end{equation}} 

\def\bsplit{\begin{split}}
\def\nsplit{\end{split}}

\begin{document}

\title{Quantum Gravity Simulation: 
{\em \normalsize Quantum simulation with a minimum length based on the generalised uncertainty principle}}

\author{Jack Keable-Elliott} 
\affiliation{School of Mathematics and Physics, University of Portsmouth, Portsmouth PO1 3QL, UK}

\author{David J. Bacon}
\affiliation{Institute of Cosmology and Gravitation, University of Portsmouth, Burnaby Road, Portsmouth PO1 3FX, UK}

\author{Andrew Burbanks}
\affiliation{School of Mathematics and Physics, University of Portsmouth, Portsmouth PO1 3QL, UK}


\author{Jaewoo Joo} 
\email{corrresponding author: \href{mailto:jaewoojoo@port.ac.uk}{jaewoojoo@port.ac.uk} }
\affiliation{School of Mathematics and Physics, University of Portsmouth, Portsmouth PO1 3QL, UK}

\date{\today}

\begin{abstract}
We present a recipe for simulating one-dimensional quantum systems for low and high energies with $L$ qubits within the framework of first quantisation. Assuming a minimum grid spacing $\Delta L$ in the finite-difference method, the generalised uncertainty principle (GUP) is derived analytically and shows distinct properties for low- and high-energy quantum systems. In the low-energy regime, the GUP approaches to the standard Heisenberg uncertainty principle (HUP) for $ \Delta L \ll \hbar$. However, for finite $\Delta L \neq 0$, the GUP mathematically provides three different regions dependent on the average momentum and suggests that a wavefunction can exhibit a single-point localisation at the finite momentum uncertainty due to lack of grid resolution. We then derive a new expression for the high-energy momentum and focus on two specific cases relevant to relativistic aspects. First, if the HUP is relaxed and the high-energy momentum is matched to the special-relativistic one, the resulting GUP predicts the existence of a minimum length with non-zero mass for high energy despite the continuous limit $\Delta L = 0$. Second, enforcing consistency with the HUP allows the recovery of a canonical high-energy representation with no minimum length. But this requirement brings incompatibility with the special-relativistic momentum and suggests a modified energy-momentum equation. Therefore, we believe that the proposed momentum formulation provides a novel pathway to investigate quantum gravity phenomena for high energy using quantum simulation tools.
\end{abstract}

\maketitle

\emph{Introduction.} -- Research in quantum gravity aims to unify general relativity with quantum mechanics. Even though both theories are remarkably successful within their respective domains, it is known that they are mathematically incompatible. Relativity provides an accurate description of gravitational phenomena at macroscopic scales, whereas quantum mechanics governs the behaviour of microscopic systems. A widely discussed feature in theoretical quantum gravity implies the existence of a minimum length scale, commonly associated with the Planck length \cite{Bosso2023,UP_review,Adler10,Kempf95,Wagner23}. Such a fundamental length is expected to regulate short-distance physics and to modify the spacetime structure at high energies.

A common phenomenological framework incorporating the concept of the minimum length is the generalised uncertainty principle (GUP), which deforms the Heisenberg uncertainty principle (HUP) and suggests the modified behaviour of quantum systems in the high-energy regime. Although numerous formulations of the GUP have been investigated theoretically \cite{Bosso2023,YJPark06,Veneziano89, Maggiore93, Garay95,GUP_Wagner2021}, direct experimental access and simulation capacity to quantum gravitational effects remains extremely challenging \cite{Ghifari25, Ong18, Ali13, MSpaper00, SBpaper00, NewquantumG25, Petruzziello21-1, Plenio20}. This limitation motivates the exploration of alternative approaches for probing quantum gravity phenomena.

In recent years, quantum simulation has emerged as a powerful platform for exploring challenging research areas in physics \cite{IBMQ,Quantinuum,Fischer26,Meglio24}. Within the first-quantised framework, the wavefunction can be represented on a discretised spatial grid, which intrinsically introduces a minimum grid length $\Delta L_d$ determined by the number of qubits used to encode a $d$-dimensional ($d$-D) quantum system. For a 1-D system, the minimum grid is given by $\Delta L = 1/2^L$, where $L$ denotes the number of qubits employed in the quantum simulation. The corresponding spatial wavefunction can be written as 
\begin{eqnarray}
\ket{\psi(x)} = \sum_{j=1}^{2^L} c_j \ket{j},
\label{eq:psi_discrete}
\end{eqnarray}
where state $\ket{j}$ corresponds to the position $x_j = j\,\Delta L$ for $0 < x \leq 1$ and the normalisation is $\sum_{j=1}^{2^L} |c_j|^2 = 1$ with probability amplitudes $c_j$ \cite{JJ2020,JJ2021,JJ2023,JJ2025}. Note that the unit length of the quantum simulation is in general expected to be a scaled length for the size of quantum systems. This qubit representation implies that only $L$ qubits are required to express $2^L$ probability amplitudes in $\ket{\psi(x)}$. In the limit $\Delta L \to 0$ ($L \to \infty$), the discretised description approaches a continuous-variable wavefunction. 

We here develop a quantum simulation method in the first-quantised description of 1-D quantum systems with fixed $\Delta L \neq 0$ and investigate how the GUP emerges from finite-difference representations of position and momentum operators. We distinguish low- and high-energy regimes and analyse how different choices of momentum operators influence the resulting GUP. A primary question would be how many qubits are required to describe high-energy systems with sufficient accuracy for quantum gravity study. Since this method naturally incorporates $\Delta L \neq 0$, it provides a framework to systematically study GUP and minimum-length effects. Our approach therefore clarifies the interplay among discretisation with $\Delta L$, high-energy momentum, and minimum-length aspects, and establishes a novel pathway for exploring quantum gravity phenomena using quantum simulation tools.

\emph{First-quantised quantum operators.} -- Since $\ket{\psi (x)}$ is given with respect to a 1-D space, we may define the discretised version of the position operator $\hat{x}$ and the momentum operator $\hat{p}_x$ for a low-energy quantum system. In particular, the discrete quantum operators are represented with fixed grid size $\Delta L$ for position and momentum based on the central-difference method such as
\begin{eqnarray}
    \hat{x} &=& \Delta L \sum_{l=1}^{2^L} l \ket{l} \bra{l}\, , ~{\rm and} ~~
	\hat{p}_{x} =  -\frac{i\hbar}{2 \Delta L} \left( \hat{A} - \hat{A}^{T} \right) \, , ~\label{eq:p03}
\end{eqnarray}
where $\hat{A}$ represents a (non-Hermitian) translation operator written by $\sum_{k=1}^{2^L -1} \ket{k} \bra{k+1}$, and $\hat{A}^{T}$ denotes its transposed operator with the reduced Planck constant $\hbar$. 
From the definitions, we find the commutation relations between $\hat{x}$ and translation operators such as 
$ [\hat x,\hat A]= - \Delta L\,\hat A, ~~{\rm and}~~[\hat x,\hat A^T]= \Delta L\,\hat A^T$.

For the squared momentum operator, we do not recycle the form of Eq.~(\ref{eq:p03}) but use the first-order central-difference expression known as a Laplacian operator in the finite difference method \cite{JJ2020, JJ2021, JJ2025,Jizba10}
\begin{eqnarray}    
\hat{p}_{x}^2 & = & -\frac{ \hbar^2 }{ (\Delta L)^2 } \Big( \hat{A} + \hat{A}^{T} - 2 \hat{I} \Big), ~~~ \label{eq:pSq02}
\end{eqnarray}
for identity operator $\hat{I}$. Note that $\hat{x}$, $\hat{p}_{x} $, and $\hat{p}_{x}^2$ are all Hermitian operators. It is crucial to define the squared momentum operator in the Laplacian operator expression for low energy because $\hat{p}_{x}^2$ corresponds to the standard second derivative with grid length $\Delta L$ while the algebraic square of the first-order central momentum operator $\left( \hat{p}_{x} \right)^2$ implies the second derivative operator with grid size $2 \Delta L$. For example, the energy spectrum of a quantum harmonic system for low energy agrees with the expression of $\left<\hat{p}^2_{x}\right>$ instead of $\left<(\hat{p}_{x})^2\right>$. Thus, the definitions of both $\hat{p}_{x}$ and $\hat{p}_{x}^2$ with same $\Delta L$ well describe low-energy quantum systems with the same order accuracy. 

\emph{Low-energy GUP with grid length $\Delta L$.} --
Following the definition of squared momentum uncertainty, $ (\delta p_{x})^2 = \left< \hat{p}_{x}^2 \right> - \left< \hat{p}_{x} \right>^2 $ as the variance of $\hat{p}_{x}$, we can derive the first-order commutation relation of $\hat{x}$ and $\hat{p}_x$ in terms of $\hat{A}$ and $\hat{A}^{T}$ and its expectation value,
\begin{eqnarray}
       \left[ \hat{x},\hat{p}_{x}  \right] &=& \frac{i\hbar}{2} \left( \hat{A} + \hat{A}^{T} \right) = i \hbar \left(  \hat{I}  -  \frac{(\Delta L)^2 }{ 2 \hbar^2 }  \hat{p}_{x}^2 \right) ,~~~~ \label{low_commutation01} \\
       \left| \left< i \left[ \hat{x},\hat{p}_{x}  \right] \right> \right|
       &=& \hbar\left| 1 -  \frac{(\Delta L)^2 }{ 2 \hbar^2 }  \left( (\delta p_{x})^2 + \left< \hat{p}_{x} \right>^2 \right) \right|. \label{p_square01}
\end{eqnarray}
The condition $\left<\hat{p}_{x}^2 \right> = 0$ might not be valid for this case since the squared momentum uncertainty cannot be negative (i.e., $ (\delta p_{x})^2 = - \left< \hat{p}_{x} \right>^2 $) since $\hat{p}_{x}$ is Hermitian. Thus, $\left<\hat{p}_{x}^2 \right> $ always becomes positive unless $\left< \hat{p}_{x} \right> = 0$ (see the {\bf Supplementary Material (SM), SM1}). For the continuous 1-D coordinate ($\Delta L \rightarrow 0$), the standard HUP is given by $\delta x\, \delta p_{x} \ge {\left|\langle i [\hat{x},\hat{p}_{x}] \rangle \right|}/{2} ={\hbar}/{2} $ \cite{HR_UR}.

\begin{figure} [t]
\centering
\includegraphics[width=0.48\textwidth,trim=0cm 0cm 0cm 0.5cm]{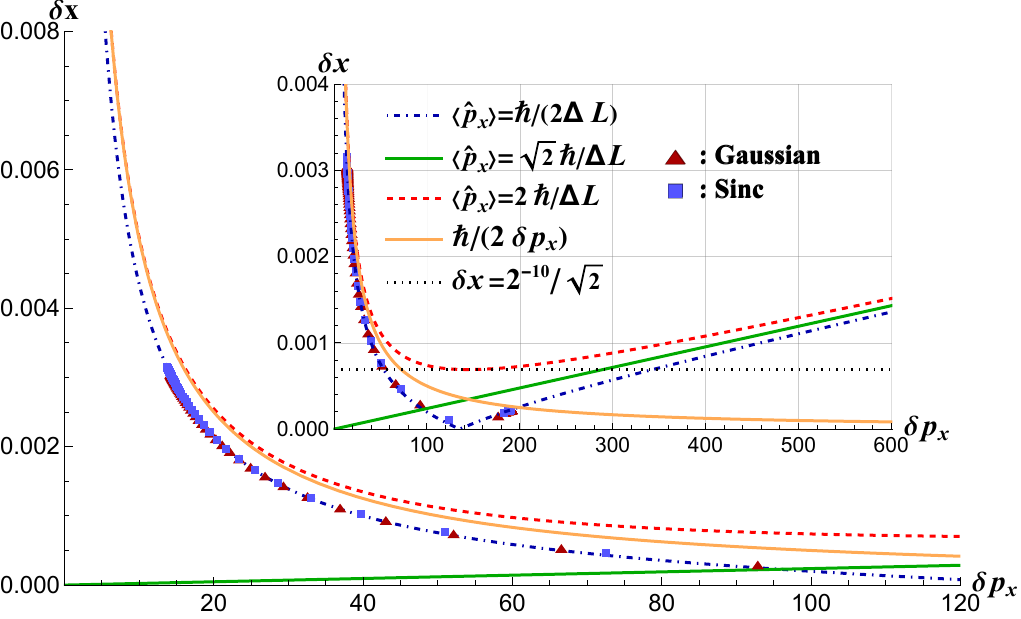}
\caption{Low-energy GUP curves from Eqs.~(\ref{LowE_uncertainty03}) to (\ref{LowE_uncertainty05}) with $\Delta L = 2^{-10}$ and $\hbar = 0.1$. The marked data are the results of numerical quantum simulation with $\left< \hat{p}_x \right> = \hbar /(2 \Delta L) \approx 51.2$ under Eq.~(\ref{LowE_uncertainty03}) for Gaussian- and sinc-like wavefunctions. The minimum length $\left( \delta x\right)_{min} = 2^{-10} / \sqrt{2}$ appears at $\left< \hat{p}_{x} \right> = 204.8$ in the dashed line while the yellow one is the HUP.}
\label{fig:01}
\end{figure}

Under the conditions $\left<\hat{p}_{x}^2 \right> > 0$ and $\Delta L \neq 0$, the low-energy GUP and its position uncertainty $\delta x$ satisfy the inequalities for $\delta p_{x} >0$ such as 
\begin{eqnarray}
&& \delta x \, \delta p_{x} \ge \frac{\hbar}{2} \left| 1 -  \frac{(\Delta L)^2 }{ 2 \hbar^2 }  \left( (\delta p_{x})^2 + \left< \hat{p}_{x}\right>^2 \right) \right| , \label{eq:LowE_uncertainty01} \\
 && \delta x \,  \ge \frac{\hbar}{2} \left| \left( 1- \frac{(\Delta L)^2 }{ 2 \hbar^2 } \left< \hat{p}_{x} \right>^2 \right) \frac{1}{\delta p_{x}}  -  \frac{(\Delta L)^2 }{ 2 \hbar^2 }  \delta p_{x} \right| . ~~~~~~~ \label{eq:LowE_uncertainty02}
\end{eqnarray}
This position uncertainty $\delta x$ in principle provides three different regions dependent on $\left< \hat{p}_{x} \right>$ given by
\begin{eqnarray}
&&  \left| \left< \hat{p}_{x} \right> \right| <  \frac{\sqrt{2} \hbar}{\Delta L} \rightarrow  \delta x \,  \ge 0~ , \label{LowE_uncertainty03} \\
&&  \left| \left< \hat{p}_{x} \right> \right| = \frac{\sqrt{2} \hbar}{\Delta L} \rightarrow  \delta x =  \frac{(\Delta L)^2 }{ 4 \hbar }  \delta p_{x}\, , 
        \label{LowE_uncertainty04} \\
&&  \left| \left< \hat{p}_{x} \right> \right| > \frac{\sqrt{2} \hbar}{\Delta L} \rightarrow  \delta x \,  \ge (\delta x)_{min} = \frac{(\Delta L)^2 }{ 2 \hbar }\,\delta p_{x}^{min} ,~~~~~~ \label{LowE_uncertainty05}
\end{eqnarray}
for $\delta p_{x}^{min} = \sqrt{ \left< \hat{p}_{x} \right>^2 -  2 \hbar^2 /(\Delta L)^2 }$. One of the extreme cases is $ \left< \hat{p}_{x} \right> =0$ under Eq.~(\ref{LowE_uncertainty03}) for which the position uncertainty in Eq.~(\ref{eq:LowE_uncertainty02}) becomes \cite{Jizba10}
\begin{eqnarray}
&& \delta x \,  \ge \frac{\hbar}{2 \, \delta p_{x} } \Big| 1  -  \frac{(\Delta L)^2 }{ 2 \hbar^2 } \left( \delta p_{x} \right)^2 \Big| \, . 
\label{eq:LowE_uncertainty06}
\end{eqnarray}
In the case $ \Delta L / \hbar \to 0 $, Eq.~(\ref{eq:LowE_uncertainty06}) approaches to the HUP (see the yellow solid line in Fig.~\ref{fig:01}). Moreover, a localised wavefunction with small $\Delta L \neq 0$ may be allowed to have zero position uncertainty (e.g., $\delta x =0$) at $\delta p_{x} =  \sqrt{ \left< \hat{p}_{x}^2 \right>} = \sqrt{2}{\hbar}/{\Delta L}$ with $ \left< \hat{p}_{x} \right> =0$ due to $\left< \left[ \hat{x},\hat{p}_{x}  \right] \right> =0$. For $\delta p_{x} \gg \sqrt{2}{\hbar}/{\Delta L}$ and $ \left< \hat{p}_{x} \right> =0$, the value of $\delta x$ is mainly proportional to that of $\delta p_{x}$ in Eq.~(\ref{eq:LowE_uncertainty06}).

\emph{Low-energy GUP examples.} -- If we choose $\Delta L = 2^{-10}$ and $\hbar = 0.1$ with three non-zero momentum cases, the GUP in Eq.~(\ref{eq:LowE_uncertainty02}) presents three coloured bounds as shown in Fig.~\ref{fig:01}. It is important to mention that these curves look similar to Fig.~1 in Ref.~\cite{UP_review} due to the $\left< \hat{p}_{x}^2 \right> $ contribution in the GUP form, however, the main difference may be caused by the fact that Eq.~(\ref{eq:LowE_uncertainty01}) has the negative sign inside the absolute value in contrast to a positive sign in Ref.~\cite{UP_review}. In Fig.~\ref{fig:01}, blue dot-dashed lines show the case of $\delta x =0$ with fixed momentum $\left< \hat{p}_{x} \right> = \hbar /(2 \Delta L) = 51.2$ at $\delta p_x =  \sqrt{7/4} \, \hbar/ \Delta L  \approx 135.46$ from Eq.~(\ref{eq:LowE_uncertainty02}). Moreover, Eq.~(\ref{LowE_uncertainty05}) seems to suggest the opportunity of having the minimum length even in the low-energy regime. For fixed $\left< \hat{p}_{x} \right> = 2 \hbar /\Delta L = 204.8$, $\left( \delta x\right)_{min} = \Delta L / \sqrt{2} = 2^{-10} / \sqrt{2}$ is shown from the red dashed curve in Fig.~\ref{fig:01}. The curves from Eqs.~(\ref{LowE_uncertainty03}) and (\ref{LowE_uncertainty05}) asymptotically approach to the green straight line in Eq.~(\ref{LowE_uncertainty04}) for large $\delta p_x$. This feature may be linked with the fact that the large $\delta p_x$ in the ultraviolet region corresponds to large $\delta x$ in the infrared one in the GUP literature \cite{UP_review}.

We found that a given discretised, localised, and normalised pure wavefunction may have a maximum momentum such as $\left< \hat{p}^{max}_{x} \right> = {\hbar}/{\Delta L}$ with the low-energy expression. This implies that the localised low-momentum wavefunction could only follow the condition of Eq.~(\ref{LowE_uncertainty03}). For instance, if we prepare a discretised ansatz state $\ket{\psi (x)}$ with $\Delta L$ corresponding to $\psi (x) = \phi^f (x)\, \exp \left( i {p}_{0} \, x /\hbar \right)$, the average low-energy momentum is approximately given by $\left< \hat{p}_x \right> \simeq {p}_{0} \leq \hbar / \Delta L$. For the maximum case $\left< \hat{p}^{max}_{x} \right> = {\hbar}/{\Delta L}$, Eq.~(\ref{eq:LowE_uncertainty02}) suggests that our simulation may achieve $\delta x =0$ with $\delta p_{x}  = {\hbar}/{\Delta L} $ and consequent $\left< \hat{p}_{x}^2 \right> = 2 {\hbar^2}/{(\Delta L)^2}$, which implies that the particle has the fixed momentum and fixed momentum-uncertainty under the GUP in Eq.~(\ref{eq:LowE_uncertainty01}).

We test two important localised classes of wavefunctions: a generalised Gaussian state $\ket{\psi_g (x)}$ and a sinc-type wavefunction $\ket{\psi_s (x)}$ in 1-D grids such as
$\ket{\psi_{g/s} (x)} = {\mathcal N}_{g/s} \sum_{j=1}^{2^L} \phi^{g/s} (x_j) \, \exp \left( i\, p_0 \, x_j /\hbar \right) \ket{j}$ where $\phi^{g}  (x_j) = \exp \left(-  (x_j-x_0)^2/(4\sigma^2) \right) $ for $x_j = j \Delta L$ and $\phi^s  (x_j) = \text{sinc} \left( \left( x_j - x_0\right)/a \right) $ for small $a$ with normalisation $\mathcal N^{-2}_{g/s} = \sum_{j=1}^{2^L} | \phi^{g/s}  (x_j) |^2 $. The average momentum for both wavefunctions with ${ 0 < \Delta L \ll \hbar}$ is given by 
\begin{eqnarray}
\left< \hat{p}_x \right> = \bra{\psi_{g,s} (x)} \hat{p}_x  \ket{\psi_{g,s} (x)} \approx \frac{\hbar}{\Delta L} \sin\!\Big(\frac{p_0 \Delta L}{\hbar}\Big)
\approx {p}_{0}.~~
\end{eqnarray}
Based on the above description of wavefunctions, we demonstrate the method of quantum simulation given by the localised ansatz states with $x_0 = 1/2$ and $\hbar = 0.1$. We use localised Gaussian and sinc states with 10 qubits ($\Delta L = 2^{-10}$) with $p_0 \approx 53.6$ or 268.08, which provide $\left< \hat{p}_{x} \right> \approx {\hbar}/ (2 \Delta L)=51.2$ for both. Thus, the quantum simulation data closely follow the blue dot-dashed curves in Fig.~\ref{fig:01} (see the details in {\bf SM2 and SM3}).

Let us consider two specific examples for $\left< \hat{p}^{max}_x \right>$. For an electron with its rest mass $m_e = 9.11 \times 10^{-31}~[kg]$ in the SI-based units, the maximum low-energy momentum with a localised ansatz in principle can be $ \left< \hat{p}_x^{max} \right> = \hbar / \Delta L^{max} = m_e v^{max} \approx m_e c$ with light speed $c$ in Eq.~(\ref{LowE_uncertainty03}). Then, its grid length becomes $\Delta L^{max} = {\hbar}/{(m_e c)} \approx 3.86 \times 10^{-13}~[m]$ known as the reduced Compton length $\lambdabar = {\hbar}/(m_e c)$ \cite{Compton}. For one meter-length quantum simulation with $L\le 41$, the maximum speed of the electron is always below the light speed $c$ such that $v^{max} = \hbar / (m_e \, \Delta L) \le 2^{41} \,\hbar / m_e \approx 2.54 \times 10^8~[m/s] \approx 0.849\,c$. Since a 42-qubit quantum simulation for an electron provides the grid length $\Delta L = 2^{-42} \approx 2.27 \times 10^{-13}~[m] < \Delta L^{max}$, the low-energy quantum simulation with $L > 41$ needs to be carefully chosen with its momentum value $\left< \hat{p}_x \right> < m_e \,c$ due to the possibility that $\left< \hat{p}_x \right> = \hbar / \Delta L > m_e \, c$. 
For another particle with the Planck momentum ${p}_{pl} = \sqrt{\hbar\,c^3 / G} \approx 6.5~[kg\,m/s]$ ($G$: gravitational constant), the Planck length with ${p}_{pl}$ is estimated by $ l_{pl} = \hbar / {p}_{pl} = \sqrt{ \hbar \, G / c^3} \approx 1.616 \times 10^{-35} \,[m]$, and we need at least 116 qubits for quantum simulation to acheive $\Delta L = 2^{-116} \approx 1.2 \times 10^{-35}~[m] < l_{pl}$. Thus, for $L \ge 116$, $p_{pl} < \left< \hat{p}^{max}_x \right>$ and the system within the Planck regime may be dealt in the low-energy viewpoint under Eq.~(\ref{LowE_uncertainty03}).

\emph{Commutation relations for high-order momentum operators.} -- We now investigate the generalised commutator expression $ [\hat{x},\hat{p}_{x}^n] $ to find the high-energy momentum and its GUP. Using the first-order commutator in Eq.~(\ref{low_commutation01}), with $\hat{p}_{x}$ and $\hat{p}_{x}^2$, we achieve the generalised commutation relation with $\hat{x}$ and the high-order power of $\hat{p}_{x}$ given as 
\begin{eqnarray}
\left[ \hat{x},{\hat{p}_x}^{2q-1} \right] &=& i\hbar \Big( (2q-1) {\hat{p}_x}^{2q-2} - q \frac{(\Delta L)^2}{2\hbar^2} {\hat{p}_x}^{2q} \Big), ~~ \label{eq:general_comm01} \\
\left[ \hat x, \hat p_{x}^{2q} \right] &=& 2 i \hbar \,q\, \hat p_{x}^{2q-1} \, ,  \label{eq:general_comm02}
\end{eqnarray}
for positive integer $q$ and $\hat{p}_{x}^0 = \hat{I}$ (see the details in {\bf SM4}).

To find the high-energy GUP, we define a deformed momentum operator $\hat{p}^H_x$ and a new commutation relation with the same position operator $ \hat{x}^H \equiv \hat{x}$, which preserves the wavefunction form in the quantum simulation. Then, $\hat{p}^H_x$ is given by a polynomial function of $\hat{p}_{x}$ \cite{DeformP01,Gusson2021}
\begin{eqnarray}
    {\hat{p}}^{H}_{x} &=& \sum_{n=1}^{\infty} \kappa_{n-1} \, \hat{p}_{x}^{n} \, . \label{eq:p_H01}
\end{eqnarray}
Since $\left[\hat{x},{\hat{p}}^H_{x} \right] = \sum_{n} \kappa_{n-1} [\hat{x},\hat{p}_{x}^n]$, the high-energy GUP is 
\begin{eqnarray}
        (\delta x)\, (\delta p^H_{x}) &\ge& \frac{\left|\langle i [\hat{x},\hat{p}^H_{x}] \rangle \right|}{2} =
        \frac{\hbar}{2} \Bigg( \kappa_0 + \sum_{t=1}^{\infty}  \tau_t\,  \left< \hat{p}_{x}^{\, t} \right> \Bigg), ~~~~~~ \label{High-energy__GUP02} 
\end{eqnarray}
where $\tau_t = (t+1) \, \kappa_{t}$ for odd $t$ and $\tau_t = (t+1) \, \kappa_t -  t \, \kappa_{t-2}\, (\Delta L)^2 / (4 \hbar^2) $ for even $t$.

For $\kappa_0 = 1$, we introduce a new GUP form for high energy in Eq.~(\ref{High-energy__GUP02}) and conclude that the different forms of ${\hat{p}}^{H}_{x}$ demonstrate two distinct features in high-energy quantum systems. First, we can adjust the parameters of $\kappa_{n-1}$ to achieve consistency between the special relativistic and our high-energy momentum expressions \cite{Pedram2012}. This compatibility imposes an inevitable minimum length for high energy despite $\Delta L \to 0$. Second, it can recover the HUP from $\left<\left[\hat{x} ,\hat{p}_{x}^H\right] \right> =  i \hbar $ with appropriate parameters $\tau_t =0$ for all $t$ in Eq.~(\ref{High-energy__GUP02}). It is crucial to point out that this condition would naturally allow us to obtain the canonical expression of the high-energy momentum operator and to maintain the HUP despite the discretised expressions for $\hat{x}$ and $\hat{p}_{x}^H$. However, this approach shows the discrepancies between the special relativity expressions and the high-energy ones. 

\emph{Mass-dependent GUP adapted to special relativity.} --
We can first select the form of high-energy momentum in Eq.~(\ref{eq:p_H01}) matched with the special relativistic momentum expression \cite{Das19}. If we adopt non-relativistic and relativistic momentum expressions, $\left<\hat{p}_{x}\right> = m v$ and ${p}^{re}_x = \gamma \left<\hat{p}_{x}\right>$ with $ \gamma = (1-v^2/c^2)^{-1/2}$, the relativistic momentum expectation value is written with $ 0 \le \left< \hat{p}_x \right>/(mc) < 1$ in 
\begin{eqnarray}
{p}^{re}_x &=& \left< \hat{p}_x \right> / \sqrt{ 1 - \left(\frac{\left< \hat{p}_x \right>}{mc}\right)^2 } =  \sum_{ {\rm odd}\,n} \kappa^{re}_{n-1} \left< \hat{p}_x \right>^{n} , ~~~~
\label{Momentum-equiv03}
\end{eqnarray} 
with $\kappa_{n-1}^{re}  =  (n-1)! \left( 2^{n-1} \left( \left( \frac{n-1}{2} \right)! \right)^2 (mc)^{n-1} \right)^{-1}$.
Note that the series converges slowly to the ultimate value for $\left<\hat{p}_x\right> \lesssim mc$. Thus, inspired by Eq.~(\ref{Momentum-equiv03}), we consider the relativistic momentum operator written as  
\begin{eqnarray}
\hat{p}^{\,re}_{x} &=& {\hat{p}_x}/{\sqrt{1 - \frac{\hat{p}_{x}^2}{m^2c^2}}} = \sum_{ {\rm odd}\,n} \kappa^{re}_{n-1} \, \hat{p}_{x}^{n} \,.
\label{eq:Relative_mom02}
\end{eqnarray}  

To find the condition $\hat{p}_x^H = \hat{p}_x^{re}$, we employ a strong conjecture that the special relativistic momentum does not take into account the particle's momentum uncertainty and holds the characteristics of a point-momentum particle with fixed speed $v$. For example, a relativistic object might be dealt as a point-momentum particle from $\left< \hat{p}_{x}^2 \right> = \left< \hat{p}_{x} \right>^2 $. Although the higher-order momentum uncertainty may be simplified as $\delta p^{(n)}_x \equiv \left< \hat{p}_{x}^n \right> - \left< \hat{p}_{x} \right>^n \ge 0$, we use here the generalised conjecture that the $n$-th order momentum becomes the $n$-th power of the low-energy momentum holding $\delta p^{(n)}_x = 0$ to select the coefficients of $\hat{p}^H_x$ (see the detailed analysis in {\bf SM5}). Then, taking the expectation value of the high-energy momentum yields the series expansion of the expectation value of the low-energy momentum from the relativistic viewpoint as
\begin{eqnarray}
\left< {\hat{p}}^{H}_{x} \right> & = & \sum_{ {\rm odd}\,n} \kappa_{n-1} \left< \hat{p}_{x}^{n} \right> \approx \sum_{ {\rm odd}\,n} \kappa_{n-1}\, \left< \hat{p}_{x} \right>^{n} .
\label{Momentum-equiv01}
\end{eqnarray}  

\begin{figure} [t]
\centering
\includegraphics[width=0.48\textwidth,trim= 0cm 0cm 0cm 0cm]{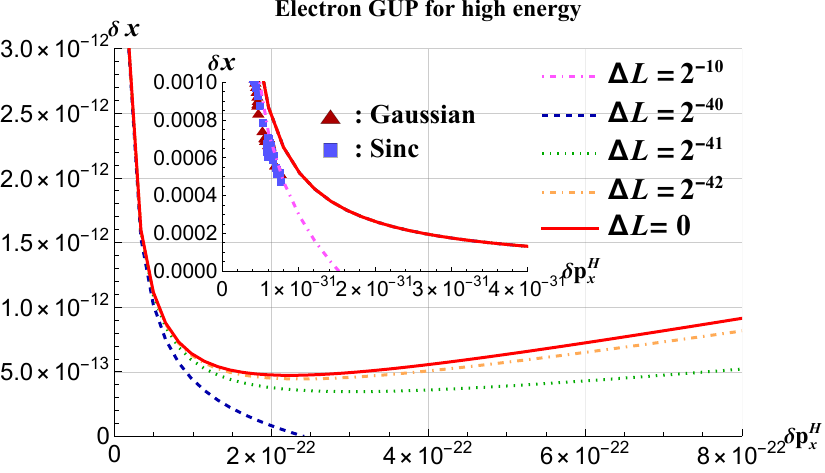}
\caption{High-energy GUP curves for an electron using different qubit numbers $L = 10, 40, 41, 42$ and $\infty$. For $\Delta L=2^{-10}$, the data for the high-energy GUP are given by the Gaussian- and sinc-type wavefunctions with a deformed high-energy momentum operator corresponding to the relativistic one.}
\label{fig:02}
\end{figure} 

We are now able to allocate the coefficients of high-energy momentum in Eq.~(\ref{eq:p_H01}) by comparing the average momentum between Eqs.~(\ref{Momentum-equiv03}) and (\ref{Momentum-equiv01}). 
For instance, $\kappa_{n-1} = \kappa^{re}_{n-1} =0$ for even $n$, $\kappa_0 = \kappa^{re}_0 =1$, $\kappa_2 = \kappa^{re}_2 = 1/(2m^2 c^2)$, $ \kappa_4 = \kappa^{re}_4 = 3/(8m^4c^4)$ and so on. This allows us to take $\left<\left[\hat{x} ,\hat{p}_{x}^H\right] \right> \equiv \left<\left[\hat{x} ,\hat{p}_{x}^{re}\right] \right> \neq  i \hbar $ due to $\tau_t \neq 0$ for even $t$ in Eq.~(\ref{High-energy__GUP02}), and the high-energy GUP becomes   
\begin{eqnarray}
        (\delta x)\, (\delta p^H_{x}) &\ge& \frac{\hbar}{2} 
        \Big( 1 +  \tau_2 \left< \hat{p}_{x}^2 \right>  +  \tau_4 \left< \hat{p}_{x}^4 \right> + \cdots \Big)\, ,~~~~~~
        \label{eq:mostHUP04} 
\end{eqnarray}
for $ \tau_2 =  \frac{3}{ 2m^2 c^2} - \frac{(\Delta L)^2 }{2 \hbar^2}$ and $ \tau_4 = \frac{15 }{ 8 m^4 c^4} - \frac{(\Delta L)^2 }{2 \hbar^2 m^2 c^2}$. A dominant feature comes from the positivity of $\tau_t$, in particular $\tau_2 >0$ with $ 0< \Delta L < \sqrt{3}\, \lambdabar$.
Moreover, since $\left(  \delta p^H_{x} \right)^2 \equiv \left< ( \hat{p}^{re}_{x} )^2 \right> - \left< \hat{p}^{re}_{x} \right>^2 \approx \left(  \delta p_{x} \right)^2 + \left< \hat{p}_{x}^4 \right>/(m^2 c^2)$, a new high-energy GUP is given with $\left< \hat{p}_{x} \right> =0$ and the continuous limit $\Delta L \to 0$ (see the details in {\bf SM6}) as
\begin{eqnarray}
        (\delta x)\, (\delta p^H_{x}) &\ge&  \frac{\hbar}{2} 
        \Big( 1 +  \tau_2  \left( \delta p^H_x \right)^2 + \cdots \big), \nonumber \\
        &=& \frac{\hbar}{2} \Big( 1 + \beta_m \, \frac{G}{\hbar c^3} \left( \delta p^H_x \right)^2 + \cdots \Big), \label{eq:mostHUP06} 
\end{eqnarray}
for the quadratic GUP-parameter $\beta_m= 3 \hbar c /(2 G m^2)$. Therefore, the high-energy GUP even with $\Delta L = 0$ naturally provides the minimum length $(\delta x)_{min}$ adapted to special relativity while the HUP holds for low energy. 

In Fig.~\ref{fig:02}, we examine the high-energy GUP with electron mass $m_e = 9.11 \times 10^{-31}~[kg]$, which determines $\beta_m \approx 8.56 \times 10^{44}$, and the range of qubit numbers show the mass-dependent GUP properties in Eq.~(\ref{eq:mostHUP06}). For instance, the quantum simulation for an electron with 40 qubits still provides $\tau_2 <0$ with the grid length $\Delta L = 2^{-40}~[m]$ while the 41-qubit quantum simulation shows the possibility to see positive minimum position uncertainty $(\delta x)_{min} > 0$ due to $\tau_2 >0$. For quantum simulation, we can utilise the low-energy wavefunction forms due to $\hat{x}^H \equiv \hat{x}$ and introduce the discretised high-energy momentum operator equivalent to the relativistic operator form in Eq.~(\ref{eq:Relative_mom02}). Inside the inset of Fig.~\ref{fig:02}, we confirm that the numerical quantum simulations for high energy with 10 qubits follow the analytical results in Eq.~(\ref{eq:mostHUP04}). Note that the slope of the asymptotic curve depends on the particle's mass. For the Planck mass $m_{pl}$, we find $\beta_m = 1.5$ with the slope of $3 \hbar / (4 m_{pl}^2 \, c^2)$. Moreover, similar-order predictions of $\beta_m$ have been discussed in the quantum gravity literature \cite{UP_review, Bosso2023, Luciano19, Das08, Astuti22, Petruzziello21}. 

\emph{HUP with discretised high-energy momentum.} --
Finally, we can hold the HUP for high energy by choosing $\tau_t = 0$ in Eq.~(\ref{High-energy__GUP02}), which has no minimum length despite the discretised momentum and position expressions. The coefficients in Eq.~(\ref{eq:p_H01}) can be set as $\kappa_0 = 1$, $\kappa_{n-1} = 0$ for even $n$ while $\kappa_{n-1}$ for odd $n\ge 3$ is survived as $\kappa_{n-1} = \frac{n-1}{4 n} \left( \frac{\Delta L}{\hbar} \right)^{2}  \kappa_{n-3} $. Based on the recurrence relation, the high-energy momentum operator is given by
\begin{eqnarray}
{\hat{p}}^{H}_{x} & = &  \hat{p}_{x} + \frac{(\Delta L)^2 }{6 \hbar^2} \hat{p}_{x}^{3} + \frac{(\Delta L)^4 }{ 30 \hbar^4 } \hat{p}_{x}^{5} + \cdots \, ,
\label{eq:HighEMom02}
\end{eqnarray} and we successfully reveal the canonical form of the commutation relation $\left< [\hat{x} ,\hat{p}_{x}^H ] \right> =  i \hbar $ for high energy. Therefore, the HUP $\delta x \, \delta p^H_{x} \ge {\hbar}/{2} $ holds even with $\Delta L \neq 0$ and we conclude no minimum length for high energy.

We now seek the possible conditions under which the high-energy momentum limit in Eq.~(\ref{eq:HighEMom02}) aligns with the special-relativistic aspects in Eq.~(\ref{eq:Relative_mom02}). When we first choose $\kappa_0 = \kappa_0^{re} = 1$ and $\kappa_2 = \kappa_2^{re}$ for the first perturbed term, the condition $\tau_2 = 0$ implies that the grid length has to be determined by $\Delta L = \sqrt{3} \hbar / (m c) = \sqrt{3}\, \lambdabar$ where the number of qubits $L$ determines the grid length $\Delta L$. However, despite the agreement of the first two terms, the rest of the terms are inevitably mismatched due to the fixed grid length. For instance, we find $\kappa_4 = (\Delta L)^4 / (30 \hbar^4) = 3 / (10 \, m^4 c^4)$ in Eq.~(\ref{eq:HighEMom02}) and thus $\kappa_4^{re} -  \kappa_4 = 3 / ( 40 m^4 c^4)$. Therefore, the high-energy momentum form $\hat{p}^{H}_{x}$ in Eq.~(\ref{eq:HighEMom02}) shows clear disagreement with $\hat{p}^{re}_x$ in Eq.~(\ref{eq:Relative_mom02}) although the high-energy HUP holds.

\emph{Conclusion and outlook.} -- In conclusion, we establish a quantum simulation framework that naturally incorporates the finite grid length $\Delta L$ to explore new types of GUPs across both low- and high-energy regimes. The results reveal novel features such as the region-dependent GUP and the possibility of vanishing position uncertainty at low energy and mass-dependent GUP with the existence of $(\delta x)_{min}$ at high energy. Two distinct high-energy momentum forms show a trade-off between consistency with special relativity and preservation of the HUP. 

This approach may offer a new pathway for investigating quantum gravity phenomena on quantum computing platforms. For example, the quantum simulation with 41 qubits may well demonstrate both low- and high-energy electron cases aligned with the special relativity. For example, in the case that the high-energy HUP holds with $\Delta L \neq 0$ in Eq.~(\ref{eq:HighEMom02}), the energy-momentum formula for high energy is
$( {\cal E}^{H}_{tot} )^2 =  m^2 c^4 + \left< {\hat{p}}^{H}_{x} \right>^2 c^2$ and shows an unavoidable discrepancy from the special relativistic one $( {\cal E}^{re}_{tot} )^2 =  m^2 c^4 + \left< {\hat{p}}^{re}_{x} \right>^2 c^2$. In particular, $( {\cal E}^{H}_{tot} )^2$ contains extra terms, which could be linked to results in a modified dispersion relation \cite{Das22}. Therefore, the grid-GUP framework with many-qubit quantum simulation in 2-D and 3-D can be performed experimentally on a quantum computer to examine results in quantum gravity.

\emph{Acknowledgements} -- JJ acknowledges T. P. Spiller, H. Kwon, K. H. Yee, Y.-W. Kim, and Y.-J. Park for useful discussion, and support from the Institute for Information \& Communications Technology Promotion (IITP) grant funded by the Korea government (MSIP) (No. 2019-000003). JK was supported by a PhD bursary from the School of Mathematics and Physics by REF2021 Quality-related Research funding.

{
\widetext
\section{Supplementary Material}

\section{SM1: Positivity of $\langle \hat p_x^2 \rangle$}
\label{Supplement01}
Let us start by defining a general form for the first-quantised localised wavefunction
\begin{eqnarray}
    \ket{\psi} = \sum_{j=1}^{2^L} c_j \ket{j},
\end{eqnarray}

with $\sum_{j=1}^{2^L} |c_j|^2 = 1$. As such, we say that our wavefunction is well localised as we are confining it to a relatively narrow region of space. The squared momentum operator for the low-energy case is defined by
\begin{eqnarray}
    {\hat{p}_x}^2 = -\frac{\hbar^2}{(\Delta L)^2} \left( \hat{A} + \hat{A}^T - 2\hat{I} \right),
\end{eqnarray}
for the shifting operators $\hat{A} = \sum_{j=1}^{2^L-1} \ket{k}\bra{k+1}$ and its transpose opearator $\hat{A}^T$. These operators are defined such that $\hat{A} \ket{k+1} = \ket{k}$ and $\hat{A}^T \ket{k} = \ket{k+1}$, so the expectation of ${\hat{p}_x}^2$ is 
\begin{eqnarray}
    \langle {\hat{p}_x}^2 \rangle = \bra{\psi} {\hat{p}_x}^2 \ket{\psi}  &=&  - \frac{\hbar^2}{(\Delta L)^2} \left[ \sum_{j=1}^{2^L - 1}   c_{j}^* c_{j+1} + \sum_{j=1}^{2^L - 1} c_{j+1}^* c_j   - 2 \right].
\end{eqnarray}

Noting that 
\begin{eqnarray}
\sum_{j=1}^{2^L - 1} | c_{j} - c_{j+1} |^2 = \sum_{j=1}^{2^L - 1} 
    \left( |c_{j}|^2 - c_{j}^* c_{j+1} - c_{j+1}^* c_{j} + |c_{j+1}|^2 \right),
\end{eqnarray}
and 
\begin{eqnarray}
    \sum_{j=1}^{2^L - 1} \left( c_{j-1}^* c_j + c_{j+1}^* c_j \right) &=&  2 - \sum_{j=1}^{2^L - 1} | c_{j} - c_{j+1} |^2 - |c_{1} |^2 - |c_{2^L}|^2.
\end{eqnarray}
Thus, we may say that
\begin{eqnarray}
    \langle {\hat{p}_x}^2 \rangle &=& \frac{\hbar^2}{(\Delta L)^2} \left( \sum_{j=1}^{2^L - 1} | c_{j} - c_{j+1} |^2 + |c_{1} |^2 + |c_{2^L}|^2 \right)
    \ge 0\, .
\end{eqnarray}
Note that the equality above only holds when $c_j = 0$ for all values of $j$. Therefore, $\langle \hat p_x^2 \rangle >0$ if $\ket{\psi}$ is a localised wavefunction.

\section{SM2: Momentum expectation value for two localised wavefunctions}
\label{Append:LocalWavefunction01}
\subsection{Generalised Form}
The general form for a localised wavefunction is
\begin{eqnarray}
    \ket{\psi (x)} = \mathcal{N} \sum_{j=1}^{2^L} \phi^f(x_j) \, \exp \left( \frac{i p_0 x_j}{\hbar} \right) \ket{j},
\end{eqnarray}
with $x_j = j\Delta L$ and the normalisation condition
\begin{eqnarray}
    \mathcal{N}^{-2} = \sum_{j=1}^{2^L} \left| \phi^f (x_j) \right|^2.
\end{eqnarray}
Applying the momentum operator to the wavefunction gives
\begin{eqnarray}
    \hat{p}_x \ket{\psi (x)} &=& -\frac{i\hbar}{2\Delta L} \mathcal{N} \sum_{j=1}^{2^L} \phi^f (x_j) \exp \left( \frac{i p_0 x_j}{\hbar} \right) \left( \hat{A} - \hat{A}^T \right) \ket{j}, \\
        &=& -\frac{i\hbar}{2\Delta L} \mathcal{N}^2  \sum_{j=1}^{2^L - 1} \bigg( \phi^* (x_{j}) \phi (x_{j+1}) \exp \left( -\frac{i p_0 x_{j}}{\hbar} \right) \exp \left( \frac{i p_0 x_{j+1}}{\hbar} \right)  \nonumber \\ 
        &&  ~~~~~~~~~~~~~~~~~~~~~~
        - \phi^*(x_{j+1}) \phi(x_j) \exp \left( -\frac{i p_0 x_{j+1}}{\hbar} \right) \exp \left( \frac{i p_0 x_j}{\hbar} \right) \bigg).
\end{eqnarray}
Using our definition of $x_j$, we may say that this is
\begin{eqnarray}
    \begin{aligned}
        \langle \hat{p}_x \rangle = -\frac{i\hbar}{2\Delta L} \mathcal{N}^2 &  \sum_{j=1}^{2^L - 1} \left(  \phi^*(x_{j}) \phi(x_{j+1}) \exp \left( \frac{i p_0 \Delta L}{\hbar} \right) -  \phi^*(x_{j+1}) \phi(x_j) \exp \left( -\frac{i p_0 \Delta L}{\hbar} \right) \right).
    \end{aligned}
\end{eqnarray}

\subsection{Gaussian Wavefunction}

For our Gaussian wavefuncion, we define
\begin{eqnarray}
    \phi^{g}(x_j) = \exp \left( -\frac{(x_j - x_0)^2}{4 \sigma^2} \right),
\end{eqnarray}
 with 
\begin{eqnarray}
    \mathcal{N}^{-2}_g = \sum_{j=1}^{2^L} \exp \left( -\frac{(x_j - x_0)^2}{2 \sigma^2} \right).
\end{eqnarray}
Then, the expectation is
\begin{eqnarray}
    \begin{aligned}
        \langle \hat{p}_x \rangle_g = -\frac{i\hbar}{2\Delta L} \mathcal{N}_{g}^2 \sum_{j=1}^{2^L - 1} &\left( \exp \left( -\frac{(x_j - \Delta L - x_0)^2}{4 \sigma^2} \right) \exp \left( -\frac{(x_j - x_0)^2}{4 \sigma^2} \right) \exp \left( \frac{i p_0 \Delta L}{\hbar} \right) \right. \\ &\left. - \exp \left( -\frac{(x_j + \Delta L - x_0)^2}{4 \sigma^2} \right) \exp \left( -\frac{(x_j - x_0)^2}{4 \sigma^2} \right) \exp \left( -\frac{i p_0 \Delta L}{\hbar} \right) \right).
    \end{aligned}
\end{eqnarray}

Using the fact that $\exp \left( -\frac{(x_j \pm \Delta L - x_0)^2}{4 \sigma^2} \right) = \exp \left( -\frac{(x_j - x_0)^2}{4 \sigma^2} \right) \exp \left( \frac{\mp \Delta L (x_j - x_0)}{2 \sigma^2} \right) \exp \left( -\frac{(\Delta L)^2}{4 \sigma^2} \right)$, we may say
\begin{eqnarray}
    \begin{aligned}
        \langle \hat{p}_x \rangle_g = \frac{\hbar}{\Delta L} \mathcal{N}_{g}^2 \exp \left( -\frac{(\Delta L)^2}{4 \sigma^2} \right) &\Bigg[ \sin \left( \frac{p_0 \Delta L}{\hbar} \right) \sum_{j=1}^{2^L - 1} \exp \left( -\frac{(x_j - x_0)^2}{2 \sigma^2} \right) \text{cosh} \left( \frac{\Delta L (x_j - x_0)}{2 \sigma^2} \right) \\ &-i \cos \left( \frac{p_0 \Delta L}{\hbar} \right) \sum_{j=1}^{2^L - 1} \exp \left( -\frac{(x_j - x_0)^2}{2 \sigma^2} \right) \text{sinh} \left( \frac{\Delta L (x_j - x_0)}{2 \sigma^2} \right) \Bigg].
    \end{aligned}
\end{eqnarray}

The imaginary sum vanishes since all values of $j$ from $j=1$ to $j=2^L - 1$ cancel perfectly, (except for the central term $j = 2^{L-1}$ where the term itself equals zero). Thus we have
\begin{eqnarray}
    \begin{aligned}
        \langle \hat{p}_x \rangle_g = \frac{\hbar}{\Delta L} \mathcal{N}_{g}^2 \exp \left( -\frac{(\Delta L)^2}{4 \sigma^2} \right) \sin \left( \frac{p_0 \Delta L}{\hbar} \right) \sum_{j=1}^{2^L - 1} \exp \left( -\frac{(x_j - x_0)^2}{2 \sigma^2} \right) \text{cosh} \left( \frac{\Delta L (x_j - x_0)}{2 \sigma^2} \right).
    \end{aligned}
\end{eqnarray}
Assuming that $\sigma \ll \left|1-x_0\right|/\sqrt{2}$, we may say that
\begin{eqnarray}
    \langle \hat{p}_x \rangle_g \approx \frac{\hbar}{\Delta L} \mathcal{N}_{g}^2 \exp \left( -\frac{(\Delta L)^2}{4 \sigma^2} \right) \sin \left( \frac{p_0 \Delta L}{\hbar} \right) \left[ \sum_{j=1}^{2^L} \exp \left( -\frac{(x_j - x_0)^2}{2 \sigma^2} \right) \text{cosh} \left( \frac{\Delta L (x_j - x_0)}{2 \sigma^2} \right) \right].
\end{eqnarray}
Moreover, if $\sigma \gg \sqrt{(\Delta L/2)}$, we may say that $\text{cosh} \left( \frac{\Delta L (x_j - x_0)}{2 \sigma^2} \right) \approx 1 + \frac{(\Delta L)^2 (x_j - x_0)^2}{8\sigma^4} \approx 1$.
Therefore, we obtain
\begin{eqnarray}
    \langle \hat{p}_x \rangle_g \approx \frac{\hbar}{\Delta L} \sin \left( \frac{p_0 \Delta L}{\hbar} \right).
\end{eqnarray}

One important note here is that, in the region $\sqrt{(\Delta L/2)} \ll \sigma \ll \left|1-x_0\right|/\sqrt{2}$, we say that $\langle \hat{p}_x \rangle_g \approx p_0$ (when $\Delta L \ll \hbar/p_0$). However, when $\sigma \lesssim \sqrt{(\Delta L/2)}$, the $\sigma$-dependent terms become more dominant, and $\langle \hat{p}_x \rangle_g < \frac{\hbar}{\Delta L} \sin \left( \frac{p_0 \Delta L}{\hbar} \right)$. So, we find that, when $\sigma$ is varied, the expectation value $\langle \hat{p}_x \rangle_g$ also varies. The choice of $p_0$ is also important here. For example, when $\sigma = \sqrt{(\Delta L/2)}$, the expectation value $\langle \hat{p}_x \rangle_g \approx 0.9998p_0$ when $p_0 = 0.01$, while the expectation value $\langle \hat{p}_x \rangle_g \approx 0.9605p_0$ when $p_0 = 50$. As such, it is crucial that the values of $\sigma$ and $p_0$ are both chosen sensibly.

\subsection{Sinc Wavefunction}
For the sinc wavefuncion, we define
\begin{eqnarray}
    \phi^{s}(x_j) = \text{sinc} \left( \frac{x_j - x_0}{a} \right),
\end{eqnarray}
and the expectation is
\begin{eqnarray}
    \begin{aligned}
        \langle \hat{p}_x \rangle_s = -\frac{i\hbar}{2\Delta L} \mathcal{N}_{s}^2 \sum_{j=1}^{2^L - 1} &\left( \text{sinc} \left( \frac{x_j - \Delta L - x_0}{a} \right) \text{sinc} \left( \frac{x_j - x_0}{a} \right) \exp \left( \frac{i p_0 \Delta L}{\hbar} \right) \right. \\ &\left. - \text{sinc} \left( \frac{x_j + \Delta L - x_0}{a} \right) \text{sinc} \left( \frac{x_j - x_0}{a} \right) \exp \left( -\frac{i p_0 \Delta L}{\hbar} \right) \right).
    \end{aligned}
\end{eqnarray}
If we expand $\text{sinc} \left( \frac{x_j \pm \Delta L - x_0}{a} \right)$ and use some useful formulas such as $\left( \frac{x_j - x_0}{x_j \pm \Delta L - x_0} \right) = 1 \mp \left( \frac{\Delta L}{x_j \pm \Delta L - x_0} \right)$, 
\begin{eqnarray}
        \text{sinc} \left( \frac{x_j \pm \Delta L - x_0}{a} \right) &=&  \text{sinc} \left( \frac{x_j - x_0}{a} \right) \cos \left( \frac{\Delta L}{a} \right) \nonumber \\ 
        &&\left( \frac{\Delta L}{x_j \pm \Delta L - x_0} \right) \left[ \text{sinc} \left( \frac{x_j - x_0}{a} \right) \cos \left( \frac{\Delta L}{a} \right)- \cos \left( \frac{x_j - x_0}{a} \right) \text{sinc} \left( \frac{\Delta L}{a} \right) \right],
\end{eqnarray}
and grouping the exponential terms with
\begin{eqnarray}
        \left( \frac{\Delta L}{x_j - \Delta L - x_0} \right) = \Bigg[ \left( \frac{\Delta L}{x_j + \Delta L - x_0} \right) + \left( \frac{2 (\Delta L)^2}{(x_j - x_0)^2 - (\Delta L)^2} \right) \Bigg],
\end{eqnarray}
we can simplify our expression to
\begin{eqnarray}
    \begin{aligned}
        \langle \hat{p}_x \rangle_s &= -\frac{i\hbar}{2\Delta L} \mathcal{N}_{s}^2 \sum_{j=1}^{2^L - 1} \\ &\Bigg[ 2i \sin \left( \frac{p_0 \Delta L}{\hbar} \right)\text{sinc}^2 \left( \frac{x_j - x_0}{a} \right) \cos \left( \frac{\Delta L}{a} \right) \Bigg. \\ & \Bigg. +\text{sinc}^2 \left( \frac{x_j - x_0}{a} \right) \cos \left( \frac{\Delta L}{a} \right) \Big[ \left( \frac{2\Delta L}{x_j + \Delta L - x_0} \right) \cos \left( \frac{p_0 \Delta L}{\hbar} \right) + \left( \frac{2(\Delta L)^2}{(x_j -x_0)^2 - (\Delta L)^2} \right) \exp \left( \frac{i p_0 \Delta L}{\hbar} \right) \Big] \\ & -\text{sinc} \left( \frac{2(x_j - x_0)}{a} \right) \text{sinc} \left( \frac{\Delta L}{a} \right) \Big[ \left( \frac{2\Delta L}{x_j + \Delta L - x_0} \right) \cos \left( \frac{p_0 \Delta L}{\hbar} \right) + \left( \frac{2(\Delta L)^2}{(x_j -x_0)^2 - (\Delta L)^2} \right) \exp \left( \frac{i p_0 \Delta L}{\hbar} \right) \Big] \Bigg].
    \end{aligned}
\end{eqnarray}

The real and imaginary terms can be simplified to give
\begin{eqnarray}
    \begin{aligned}
        \langle \hat{p}_x \rangle_s &= \frac{\hbar}{\Delta L} \mathcal{N}_{s}^2 \Bigg[ \sin \left( \frac{p_0 \Delta L}{\hbar} \right) \cos \left( \frac{\Delta L}{a} \right) \sum_{j=1}^{2^L - 1} \text{sinc}^2 \left( \frac{x_j - x_0}{a} \right) \Bigg. \\ &\phantom{=\frac{\hbar}{\Delta L} \mathcal{N}_{s}^2}+ \sin \left( \frac{p_0 \Delta L}{\hbar} \right) \cos \left( \frac{\Delta L}{a} \right) \sum_{j=1}^{2^L - 1} \left( \frac{(\Delta L)^2}{(x_j - x_0)^2 - (\Delta L)^2} \right) \text{sinc}^2 \left( \frac{x_j - x_0}{a} \right) \\ &\phantom{=\frac{\hbar}{\Delta L} \mathcal{N}_{s}^2}- \sin \left( \frac{p_0 \Delta L}{\hbar} \right) \text{sinc} \left( \frac{\Delta L}{a} \right) \sum_{j=1}^{2^L - 1} \left( \frac{(\Delta L)^2}{(x_j - x_0)^2 - (\Delta L)^2} \right) \text{sinc} \left( \frac{2(x_j - x_0)}{a} \right) \\ &\phantom{=\frac{\hbar}{\Delta L} \mathcal{N}_{s}^2} -i \cos \left( \frac{p_0 \Delta L}{\hbar} \right) \cos \left( \frac{\Delta L}{a} \right) \sum_{j=1}^{2^L - 1} \left( \frac{\Delta L}{x_j + \Delta L - x_0} + \frac{(\Delta L)^2}{(x_j - x_0)^2 - (\Delta L)^2} \right) \text{sinc}^2 \left( \frac{x_j - x_0}{a} \right) \\ &\phantom{=\frac{\hbar}{\Delta L} \mathcal{N}_{s}^2} +i \cos \left( \frac{p_0 \Delta L}{\hbar} \right) \text{sinc} \left( \frac{\Delta L}{a} \right) \sum_{j=1}^{2^L - 1} \left( \frac{\Delta L}{x_j + \Delta L - x_0} + \frac{(\Delta L)^2}{(x_j - x_0)^2 - (\Delta L)^2} \right) \text{sinc} \left( \frac{2(x_j - x_0)}{a} \right) \Bigg].
    \end{aligned}
\end{eqnarray}

Similarly to the Gaussian case, the imaginary terms vanish too. For all values of $j$ from $j=1$ to $j=2^L - 1$, the summations perfectly cancel, leaving only a central term at $j=2^{L-1}$ which also equals zero. As a result, we are left with
\begin{eqnarray}
    \begin{aligned}
        \langle \hat{p}_x \rangle_s &= \frac{\hbar}{\Delta L} \mathcal{N}_{s}^2 \Bigg[ \sin \left( \frac{p_0 \Delta L}{\hbar} \right) \cos \left( \frac{\Delta L}{a} \right) \sum_{j=1}^{2^L - 1} \text{sinc}^2 \left( \frac{x_j - x_0}{a} \right) + \mathcal{O} \left( (\Delta L)^2 \right) \Bigg].
    \end{aligned}
\end{eqnarray}
This is equivalent to
\begin{eqnarray}
    \begin{aligned}
        \langle \hat{p}_x \rangle_s &= \frac{\hbar}{\Delta L} \mathcal{N}_{s}^2 \Bigg[ \sin \left( \frac{p_0 \Delta L}{\hbar} \right) \cos \left( \frac{\Delta L}{a} \right) \left[ \sum_{j=1}^{2^L} \text{sinc}^2 \left( \frac{x_j - x_0}{a} \right) - \text{sinc}^2 \left( \frac{x_0}{a} \right) \right] + \mathcal{O} \left( (\Delta L)^2 \right) \Bigg].
    \end{aligned}
\end{eqnarray}
In the regime in which $a \ll x_0$, we say that
\begin{eqnarray}
    \begin{aligned}
        \langle \hat{p}_x \rangle_s &\approx \frac{\hbar}{\Delta L} \mathcal{N}_{s}^2 \sin \left( \frac{p_0 \Delta L}{\hbar} \right) \cos \left( \frac{\Delta L}{a} \right) \sum_{j=1}^{2^L} \text{sinc}^2 \left( \frac{x_j - x_0}{a} \right).
    \end{aligned}
\end{eqnarray}
If we also say $\Delta L \ll a$, this simplifies to
\begin{eqnarray}
    \begin{aligned}
        \langle \hat{p}_x \rangle_s &\approx \frac{\hbar}{\Delta L} \sin \left( \frac{p_0 \Delta L}{\hbar} \right).
    \end{aligned}
\end{eqnarray}

Once again, it is important to note here that, in the region $\Delta L \ll a \ll x_0$, we say that $\langle \hat{p}_x \rangle_s \approx p_0$ (when $\Delta L \ll \hbar/p_0$). However, when $a \approx \Delta L$, the $a$-dependent terms become more dominant, and $\langle \hat{p}_x \rangle_s < \frac{\hbar}{\Delta L} \sin \left( \frac{p_0 \Delta L}{\hbar} \right)$. So, in the same way as the Gaussian, we find that when $a$ is varied, the expectation value $\langle \hat{p}_x \rangle_s$ also varies. For example, when $a = \sqrt{(\Delta L/2)}$, the expectation value $\langle \hat{p}_x \rangle_s \approx 0.9968p_0$ when $p_0 = 0.01$, while the expectation value $\langle \hat{p}_x \rangle_s \approx 0.9577p_0$ when $p_0 = 50$.

\section{SM3: Quantum Simulation in QuTip}
To simulate the low-energy GUP, the first step was to build functions corresponding to the discretised position and momentum operators. This was done using the open-source QuTiP library for python. The major benefit of using QuTiP is that the data structures are able to incorporate the properties of quantum operators and state vectors (e.g. matrices can easily be built with dimensions corresponding to those of the require states). Next, the Gaussian and sinc states were then generated in a similar way and applied to the operators in order to produce the expectation values for the position and momentum. These expectation values were then used to determine the uncertainties in momentum and position. This is shown more explicitly in Algorithm 1.
\newline
\begin{algorithm}
\caption{Low-Energy Quantum Simulation}
\begin{algorithmic}[1]
\Procedure{Generating Operators and States for Low-Energy GUP}{}
  \State $\text{Initialise } N = 10, \Delta L = 1/2^N,~ \hbar = 0.1,~ x_0 = 0.5$
  \State $\text{Initialise } p_{0},~\sigma_{range},~a_{range}$
  \State $\text{Generate } \hat{x} \text{ using } N \text{ and } \Delta L$
  \State $\text{Generate } \hat{A} \text{ and } \hat{A}^T \text{ using } N$
  \State $\text{Generate } \hat{p}_x \text{ and } {\hat{p}_x}^2 \text{ using } N \text{, } \Delta L \text{ and } \hbar$
  \State $\text{Generate } \ket{\psi}_g \text{ using } N \text{, } \Delta L \text{, } \hbar \text{, } x_0 \text{, } p_{0} \text{, } \sigma_{range}$
  \State $\text{Generate } \ket{\psi}_s \text{ using } N \text{, } \Delta L \text{, } \hbar \text{, } x_0 \text{, } p_{0} \text{, } a_{range}$
\EndProcedure

\Procedure{Calculating Expectation Values and Uncertainties for Low-Energy GUP}{}
  \State $\text{Compute expectation values for } \hat{x} \text{, } \hat{x}^2 \text{, } \hat{p}_x \text{, } {\hat{p}_x}^2 \text{ with } \ket{\psi}_g \text{ and } \ket{\psi}_s$
  \State $\text{Compute } \delta p_x \text{ using expectation values}$
  \State $\text{Compute } \delta x \text{ using Low-Energy GUP}$
\EndProcedure
\end{algorithmic}
\end{algorithm}

Fig.~\ref{SM3} shows an example of the results for the low-energy uncertainty relation. Both the Gaussian and sinc wavefunctions are individually normalised for any given value of $\sigma$ or $a$, the result of which is that, no matter how wide the wavefunction ends up being, its norm will always be 1. This means that, for wide wavefunctions where a lot of information about the tails of the distribution are lost, they are still renormalised as if the full distribution is contained within the desired region of space. This can cause complications, so it is crucial that the widths of the wavefunctions are kept narrow enough for these effects to be negligible.
\begin{figure} [ht]
\centering
\includegraphics[width=0.6\textwidth,trim=0cm 0cm 0cm 1cm]{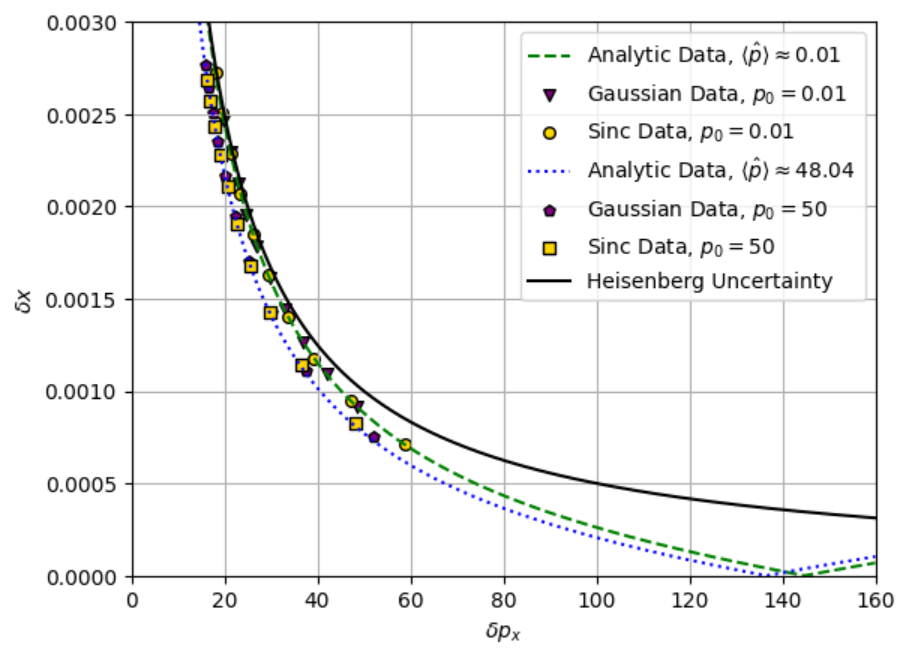}
\caption{Low-energy GUP curves using 10 qubits with $\Delta L = 2^{-10}$ and $\hbar = 0.1$. For the Gaussian wavefunction (sinc wavefunction), the choice of $\sigma$ ($a$) ranges from $0.001$ to $0.0025$ ($0.003$ to $0.01$) for $p_0 = 0.01$, and from $0.001$ to $0.005$ ($0.004$ to $0.017$) for $p_0 = 50$. } \label{SM3}
\end{figure}
A range of different values for $\sigma$ (corresponding to the Gaussian width) and $a$ (corresponding to the sinc width) were used in order to produce the results in Fig.~\ref{SM3} (the exact values stated in the figure heading). The analytical data in this figure was produced by feeding a range of test points for the momentum uncertainty into our low-energy GUP to determine the position uncertainties. Note that, for the analytical case, the momentum expectation values are 

\begin{eqnarray}
    \langle \hat{p}_x \rangle_{A} = \frac{\hbar}{\Delta L} \sin \left( \frac{p_0 \Delta L}{\hbar} \right)
\end{eqnarray}
\newline
taken from the approximation for the momentum expectation when $\sigma \gg \sqrt{\Delta L/2}$ (for Gaussian case) and  $a \gg \Delta L$ (for sinc case). As observed in the figure, both the Gaussian and sinc data points follow the analytical data well for $0 \le \delta p_x \lesssim 30$ but, past this point, the correlation starts to worsen. This happens because the values of $\sigma$/$a$ used here are starting to near the size of $\Delta L$. As a result, our assumption that $\Delta L \ll \sigma$ (or $\Delta L \ll a$) becomes naive, and thus our description for $\langle \hat{p}_x \rangle_{A}$ is unrealistic. In actuality, any terms that have a $\sigma$ or $a$ dependence become more dominant, decreasing the magnitude of $\langle \hat{p}_x \rangle$ and therefore increasing the overall magnitude of $\delta p_x$ (hence why the data points begin to sit above the analytical data). The extent to which the choice of width affects the $\langle \hat{p}_x \rangle$ calculation is displayed in TABLE 1.

\begin{table}[ht]
\centering
\begin{tabular}{ | c |c|c|c|c|c|  }
\hline
\multirow{2}{2em}{$p_0$}& \multirow{2}{6em}{Wavefunction} & \multicolumn{2}{|c|}{Minimum Widths} & \multicolumn{2}{|c|}{Maximum Widths} \\
\cline{3-6}
 &  & Width & $\langle \hat{p}_x \rangle$ & Width & $\langle \hat{p}_x \rangle$ \\
\hline \hline
\multirow{3}{2em}{0.01} & Gaussian & $\sigma = 0.001$ & 0.00888 & $\sigma = 0.0025$ & 0.00981 \\
\cline{2-6}
 & Sinc & $a = 0.003$ & 0.00835 & $a = 0.01$ & 0.00984 \\
\cline{2-6}
 & Analytic & N/A & 0.00999 & N/A & 0.00999 \\
\hline \hline
\multirow{3}{2em}{50} & Gaussian & $\sigma = 0.001$ & 42.63846 & $\sigma = 0.005$ & 47.80822 \\
\cline{2-6}
 & Sinc & $a = 0.004$ & 43.47074 & $a = 0.017$ & 47.77791 \\
\cline{2-6}
 & Analytic & N/A & 48.03673 & N/A & 48.03673 \\
\hline
\end{tabular}
\caption{The expectation values $\langle \hat{p}_x \rangle$ for the highest and lowest values of $\sigma$ and $a$, when $p_0 = 0.01$ or $p_0 = 50$. This is compared to the analytical results, all data is taken directly from data used in Fig.~\ref{SM3}.}
\end{table}

Note that the width for the analytic cases in TABLE 1 are recorded as N/A since there is no dependence on either $\sigma$ or $a$. It is clear from the results in TABLE 1 that the $\langle \hat{p}_x \rangle$ calculation obtained using the lowest values of $\sigma$ and $a$ are further from the analytic result than those obtained using the highest values. This is to be expected since, when $\sigma < \Delta L$ or $a < \Delta L$, the expectation value of the momentum is expected to decrease towards zero. These findings confirm the assumptions made regarding the reasoning for why the correlation between gaussian/sinc data and analytic result worsen as $\delta p_x$ increases.

To simulate the high-energy GUP, a new function had to be defined to describe the high-energy discretised momentum operator. The algorithm required to build this is detailed in Algorithm 2.
\begin{algorithm}
\caption{High-Energy Quantum Simulation}
\begin{algorithmic}[1]
\Procedure{Generating Operators and States for High-Energy GUP}{}
  \State $\text{Initialise } N = 10,~\Delta L = 1/2^N,~ \hbar = 1.05 \times 10^{-34},~n=15,~ m_e\,c \approx 2.73 \times 10^{-22},~ x_0 = 0.5$
  \State $\text{Initialise } p_{0},~\sigma_{range},~a_{range}$
  \State $\text{Generate } \kappa \text{ using } n \text{ and } m_e\,c$
  \State $\text{Generate } \hat{p}_x^H \text{ using } n \text{, } \hat{p}_x \text{, } {\hat{p}_x}^2 \text{ and } \kappa$
  \State $\text{Generate } \ket{\psi}_g \text{ using } N \text{, } \Delta L \text{, } \hbar \text{, } x_0 \text{, } p_{0} \text{, } \sigma_{range}$
  \State $\text{Generate } \ket{\psi}_s \text{ using } N \text{, } \Delta L \text{, } \hbar \text{, } x_0 \text{, } p_{0} \text{, } a_{range}$
\EndProcedure

\Procedure{Calculating Expectation Values and Uncertainties for High-Energy GUP}{}
  \State $\text{Compute expectation values for } \hat{x} \text{, } \hat{x}^2 \text{, } \hat{p}_x^H \text{, } (\hat{p}_x^H)^2 \text{ with } \ket{\psi}_g \text{ and } \ket{\psi}_s$
  \State $\text{Compute } \delta p_x^H \text{ using expectation values}$
  \State $\text{Compute } \delta x \text{ using High-Energy GUP}$
\EndProcedure
\end{algorithmic}
\end{algorithm}
Note that the variable $n$ used for the creation of the $\kappa$ terms defines how many $\kappa$ values are created (and therefore how many terms should be involved in the overall summation to calculate $\hat{p}_x^H$). In the same way as for the low-energy GUP, the expectation values are used to compute the high-energy uncertainties (using the high-energy GUP).

\begin{figure} [ht]
\centering
\includegraphics[width=0.55\textwidth,trim=0cm 0cm 0cm 1cm]{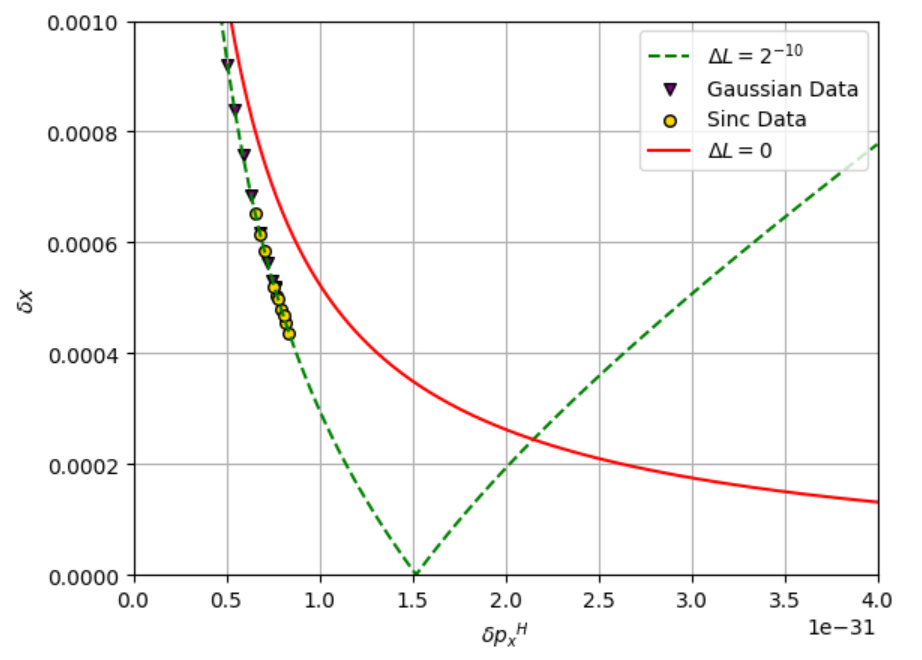}
\caption{High-energy GUP curve with $\Delta L = 2^{-10}$ [$m$] and $\Delta L = 0$ for $\langle \hat{p}_x \rangle = 0$, $\hbar = 1.05\times 10^{-34}$ [$kg\,m^2/s$], $m = m_e = 9.11 \times 10^{-31}$ [$kg$], $c = 3\times 10^8$ [$m/s$]. For the Gaussian state, the choice of $\sigma$ ranges from $0.0001$ to $0.0009$ while, for the sinc state, the choice of $a$ ranges from $0.0001$ to $0.0025$.} \label{SM4}
\end{figure}

Once again, a range of different values for $\sigma$ and $a$ were used in order to produce the results in Fig.~\ref{SM4} (the exact values stated in the figure heading). The analytical data in this figure was produced by feeding a range of test points for the momentum uncertainty into our high-energy GUP (when $\langle \hat{p}_x \rangle = 0$) to determine the position uncertainties. Note that, for the case that $\Delta L = 2^{-10}$ [$m$], there still exists a point at which the position uncertainty equals zero. However, when $\Delta L = 0$, we do not observe this effect - the position uncertainty has a non-zero minimum value in this case.

\section{SM4: Proof of the Commutator Identity}
\label{Supplement04}
\subsection{Commutation relation with even-powered momentum}

The Leibniz property of commutators states that $\left[ \hat{L},\hat{M} \hat{N} \right] = \hat{M} \left[ \hat{L}, \hat{N} \right] + \left[ \hat{L}, \hat{M} \right] \hat{N}$ for any set of operators $\hat{L}$, $\hat{M}$ and $\hat{N}$. When this is applied repeatedly, we eventually see that
\begin{eqnarray}
    \left[ \hat{L}, \hat{M}^j \right] = \sum_{k=1}^{j} \hat{M}^{k-1} \left[ \hat{L}, \hat{M} \right] \hat{M}^{j-k}.
\end{eqnarray}
Obviously, this only holds when we say that $(\hat{M})^q = \hat{M}^q$, which seems trivially true. However, in our case that $\left( \hat{p}_x \right)^2 \ne {\hat{p}_x}^2$, the above identity cannot be used. But, as long as we say that $\hat{A} \hat{A}^T = \hat{A}^T \hat{A} = \hat{I}$ (and thus $\left[ \hat{A}, \hat{A}^T \right] = 0$), we may define $\left( {\hat{p}_x}^2 \right)^q = {\hat{p}_x}^{2q}$, and thus the above identity becomes useful. We may say that
\begin{eqnarray}
    \left[ \hat{x}, {\hat{p}_x}^{2q} \right] = \sum_{k=1}^{q} \left( {\hat{p}_x}^2 \right)^{k-1} \left[ \hat{x}, {\hat{p}_x}^2 \right] \left( {\hat{p}_x}^2 \right)^{q-k}.
\end{eqnarray}

Exploiting the previously made assumption that $\left[ \hat{A}, \hat{A}^T \right] = 0$, we may simplify this to
\begin{eqnarray}
    \left[ \hat{x}, {\hat{p}_x}^{2q} \right] = q \left[ \hat{x}, {\hat{p}_x}^2 \right] \left( {\hat{p}_x}^2 \right)^{q-1}.
\end{eqnarray}
For example, the commutator $\left[ \hat{x}, {\hat{p}_x}^2 \right]$ may be found to be
\begin{eqnarray}
    \begin{aligned}
        \left[ \hat{x}, {\hat{p}_x}^2 \right] &= - \frac{\hbar^2}{(\Delta L)^2} \left[ \hat{x}, \left( \hat{A} + \hat{A}^T - 2\hat{I} \right) \right] = - \frac{\hbar^2}{(\Delta L)^2} \left( -\Delta L \hat{A} + \Delta L \hat{A}^T \right) = \frac{\hbar^2}{\Delta L} \left( \hat{A} - \hat{A}^T \right) = 2i\hbar \hat{p}_x \,,
    \end{aligned}
\end{eqnarray}
and thus we get that
\begin{eqnarray}
    \left[ \hat{x}, {\hat{p}_x}^{2q} \right] = 2i\hbar q \, {\hat{p}_x}^{2q-1}.
\end{eqnarray}
This holds for all even-powered momentum terms ($q = \{1,2,3,\dots\}$).

\subsection{Commutation relation with odd-powered momentum}
For odd-powered momentum, we say

\begin{eqnarray}
    \left[ \hat{x}, {\hat{p}_x}^{2q-1} \right] = \left[ \hat{x}, \hat{p}_x \cdot {\hat{p}_x}^{2q-2} \right] = \hat{p}_x \left[ \hat{x}, {\hat{p}_x}^{2q-2} \right] + \left[ \hat{x}, \hat{p}_x \right] {\hat{p}_x}^{2q-2}.
\end{eqnarray}
From the result in the previous subsection,
\begin{eqnarray}
    \left[ \hat{x}, {\hat{p}_x}^{2q-2} \right] = 2i\hbar(q-1) {\hat{p}_x}^{2q-3},
\end{eqnarray}
and we may also use our definition of $\left[ \hat{x}, \hat{p}_x \right]$ to say that
\begin{eqnarray}
    \begin{aligned}
        \left[ \hat{x}, {\hat{p}_x}^{2q-1} \right] &= 2(q-1)i\hbar \left( \hat{p}_x \right)^2 {\hat{p}_x}^{2q-4} + i\hbar \left( {\hat{p}_x}^{2q-2} - \frac{(\Delta L)^2}{2\hbar^2} {\hat{p}_x}^{2q} \right).
    \end{aligned}
\end{eqnarray}

According to our definitions for ${\hat{p}_x}^2$ and ${\hat{p}_x}^4$, $\left( \hat{p}_x \right)^2 = {\hat{p}_x}^2 - \frac{(\Delta L)^2}{4 \hbar^2} {\hat{p}_x}^4$, and 
\begin{eqnarray}
    \left[ \hat{x}, {\hat{p}_x}^{2q-1} \right] = 2(q-1)i\hbar \left( {\hat{p}_x}^2 - \frac{(\Delta L)^2}{4 \hbar^2} {\hat{p}_x}^4 \right) {\hat{p}_x}^{2q-4} + i\hbar \left( {\hat{p}_x}^{2q-2} - \frac{(\Delta L)^2}{2\hbar^2} {\hat{p}_x}^{2q} \right).
\end{eqnarray}
Therefore,
\begin{eqnarray}
    \left[ \hat{x}, {\hat{p}_x}^{2q-1} \right] = i\hbar \left( (2q-1) {\hat{p}_x}^{2q-2} -q \frac{(\Delta L)^2}{2\hbar^2} {\hat{p}_x}^{2q}  \right).
\end{eqnarray}
This holds for all odd-powered momentum terms ($q = \{1,2,3,\dots\}$).

\section*{SM5: Analysis for high-order momentum uncertainty}
We may say that the nth moment of the low-energy momentum is $\langle \hat{p}^n_x \rangle$, so the first moment is simply $\langle \hat{p}_x \rangle$, the momentum expectation value. From the book of G. Grimmett and D. Stirzaker, {\em Probability and Random Processes} (2009), the $n$-th order central moment for the low-energy momentum can be defined as
\begin{eqnarray}
    \delta p_x^{(n)} = \langle ( \hat{p}_x - \langle \hat{p}_x \rangle )^n \rangle.
\end{eqnarray}

\subsection{2nd Order Moment}
When $n = 2$, the result is trivially how we define the low-energy momentum uncertainty

\begin{eqnarray}
    \delta p_x^{(2)} = \langle {\hat{p}_x}^2 - 2 \hat{p}_x \langle \hat{p}_x \rangle + \langle \hat{p}_x \rangle^2  \rangle = \langle {\hat{p}_x}^2 \rangle - \langle \hat{p}_x \rangle^2 = (\delta p_x )^2.
\end{eqnarray}

This is known as the variance of the momentum, a measure of the extent by which the momentum deviates from its expectation value.

\subsection{3rd Order Moment}
When $n = 3$, the result is

\begin{eqnarray}
    \begin{aligned}
        \delta p_x^{(3)} &= \langle ( \hat{p}_x - \langle \hat{p}_x \rangle )^3 \rangle = \langle {\hat{p}_x}^3 - 3 {\hat{p}_x}^2 \langle \hat{p}_x \rangle + 3 \hat{p}_x \langle \hat{p}_x \rangle^2 - \langle \hat{p}_x \rangle^3 \rangle = \langle {\hat{p}_x}^3 \rangle - 3 \langle {\hat{p}_x}^2 \rangle \langle \hat{p}_x \rangle + 3 \langle \hat{p}_x \rangle \langle \hat{p}_x \rangle^2 - \langle \hat{p}_x \rangle^3.
    \end{aligned}
\end{eqnarray}

Using the knowledge that $\langle {\hat{p}_x}^2 \rangle = \delta p_x^{(2)} + \langle \hat{p}_x \rangle^2$, we may substitute this into the above to get

\begin{eqnarray}
    \delta p_x^{(3)} = \langle {\hat{p}_x}^3 \rangle - 3 \left( \delta p_x^{(2)} + \langle \hat{p}_x \rangle^2 \right) \langle \hat{p}_x \rangle + 3 \langle \hat{p}_x \rangle \langle \hat{p}_x \rangle^2 - \langle \hat{p}_x \rangle^3.
\end{eqnarray}

In the case that the 2nd order moment (and therefore the momentum uncertainty) is 0, this simplifies to

\begin{eqnarray}
    \delta p_x^{(3)} = \langle {\hat{p}_x}^3 \rangle - \langle \hat{p}_x \rangle^3.
\end{eqnarray}

This result, the third order moment, is related to the skewness of the momentum distribution, in particular for Gaussian states. In other words, it measures whether the distribution is biased towards higher or lower momentum values. Studying the skewness of a distribution can be helpful in determining asymmetric scattering processes or wavefunction distortions. While useful, this area of statistics is not particularly relevant to this study.

\subsection{4th Order Moment}
When $n = 4$, the result is

\begin{eqnarray}
    \begin{aligned}
        \delta p_x^{(4)} &= \langle ( \hat{p}_x - \langle \hat{p}_x \rangle )^4 \rangle =\langle {\hat{p}_x}^4 \rangle - 4 \langle {\hat{p}_x}^3 \rangle \langle \hat{p}_x \rangle + 6 \langle  {\hat{p}_x}^2 \rangle \langle \hat{p}_x \rangle^2  - 4 \langle \hat{p}_x \rangle \langle \hat{p}_x \rangle^3 + \langle \hat{p}_x \rangle^4 .
    \end{aligned}
\end{eqnarray}

Once again we may use the identity $\langle {\hat{p}_x}^2 \rangle = \delta p_x^{(2)} + \langle \hat{p}_x \rangle^2$, but this time we assume that $\delta p_x^{(2)} = 0$ so that we may say $\langle {\hat{p}_x}^3 \rangle = \delta p_x^{(3)} + \langle \hat{p}_x \rangle^3$. As a result, we can say that
\begin{eqnarray}
    \delta p_x^{(4)} = \langle {\hat{p}_x}^4 \rangle - 4 \left( \delta p_x^{(3)} + \langle \hat{p}_x \rangle^3 \right) \langle \hat{p}_x \rangle + 6 \left( \langle \hat{p}_x \rangle^2 \right) \langle \hat{p}_x \rangle^2  - 4 \langle \hat{p}_x \rangle \langle \hat{p}_x \rangle^3 + \langle \hat{p}_x \rangle^4 .
\end{eqnarray}
Thus, if we now also have that the 3rd order moment is 0, this simplifies to
\begin{eqnarray}
    \delta p_x^{(4)} = \langle {\hat{p}_x}^4 \rangle - \langle \hat{p}_x \rangle^4 .
\end{eqnarray}

The fourth order moment is known for the kurtosis of the momentum distribution. Essentially, this is a measure of how significant extreme momentum results are within the distribution. Studying the kurtosis of the distribution is helpful in finding evidence of quantum turbulence or the existence of rare (yet significant) extreme momentum components. Once again, this area of statistics is not particularly relevant to this study.

\subsection{Relevance of the $n$-th Order Moment of the Low-Energy Momentum}
We have seen that, assuming the $(n-1)$-th order moment is 0, it is fair to say that $\delta p_x^{(n)} = \langle {\hat{p}_x}^n \rangle - \langle \hat{p} \rangle^n$. As a result, it is also fair to say that, as long as $\delta p_x^{(n)} = 0$, it is true that $\langle {\hat{p}_x}^n \rangle = \langle \hat{p} \rangle^n$. What's more, we can have this be true while still accepting that higher-order moments may be non-zero (e.g. forcing $\delta p_x^{(2)} = 0$ still allows for $\delta p_x^{(3)} \ne 0$).

\section{SM6: High-Energy Momentum Uncertainty }
We may perform a Taylor expansion on our high-energy momentum operator, which is equivalent to the special relativistic one, as follows
\begin{eqnarray}
    {\hat{p}_x}^H = \hat{p}_x  \left( 1 - \frac{{\hat{p}_x}^2}{m^2 c^2} \right)^{-1/2}= \hat{p}_x + \frac{1}{2 m^2 c^2} {\hat{p}_x}^3 + \frac{3}{8 m^4 c^4} {\hat{p}_x}^5 + \dots \,.
\end{eqnarray}
This assumes that ${\hat{p}_x}^2 \ll m^2 c^2$. If we define the square of the high-energy momentum as 
\begin{eqnarray}
    ({\hat{p}_x}^H)^2 = \hat{p}_x^2 \left( 1 - \frac{{\hat{p}_x}^2}{m^2 c^2} \right)^{-1}.
\end{eqnarray}
Since there is no relevant expression for a high-energy Laplacian operator, this may similarly be expanded as
\begin{eqnarray}
    ({\hat{p}_x}^H)^2 = {\hat{p}_x}^2 + \frac{1}{m^2 c^2} {\hat{p}_x}^4 + \frac{1}{m^4 c^4} {\hat{p}_x}^6 + \dots\,.
\end{eqnarray}

If we therefore define the high-energy uncertainty in momentum as $( \delta {p_x}^H )^2 = \langle ({\hat{p}_x}^H)^2 \rangle - \langle {\hat{p}_x}^H \rangle^2$, the result is
\begin{eqnarray}
        ( \delta {p_x}^H )^2 &=& \left\langle {\hat{p}_x}^2 + \frac{1}{m^2 c^2} {\hat{p}_x}^4 + \frac{1}{m^4 c^4} {\hat{p}_x}^6 + \dots \right\rangle - \left\langle \hat{p}_x + \frac{1}{2 m^2 c^2} {\hat{p}_x}^3 + \frac{3}{8 m^4 c^4} {\hat{p}_x}^5 + \dots \right\rangle^2 , \nonumber \\ 
        &=&  ( \delta p_x )^2 + \frac{1}{m^2 c^2} \left( \langle {\hat{p}_x}^4 \rangle - \langle \hat{p}_x \rangle \langle {\hat{p}_x}^3 \rangle \right) + \dots \, .
\end{eqnarray}
We can rearrange this as an approximation of $\langle {\hat{p}_x}^2 \rangle$ and insert it into the formula for the high-energy GUP
\begin{eqnarray}
    (\delta x) (\delta {p_x}^H) &\ge& \frac{\hbar}{2} \left( 1 + \tau_2 \langle {\hat{p}_x}^2 \rangle + \tau_4 \langle {\hat{p}_x}^4 \rangle + \dots \right) , \nonumber \\ 
   &\ge& \frac{\hbar}{2} \left( 1 + \tau_2 \left( (\delta {p_x}^H)^2 + \langle \hat{p}_x \rangle^2 \right) + \left( \tau_4 - \frac{\tau_2}{m^2 c^2} \right) \langle {\hat{p}_x}^4 \rangle + \dots \right).
\end{eqnarray}
For $\tau_2 = \frac{3}{2 m^2 c^2} - \frac{(\Delta L)^2}{2\hbar^2}$ and $\tau_4 = \frac{15}{8 m^4 c^4} - \frac{(\Delta L)^2}{2\hbar^2 m^2 c^2}$, we get
\begin{eqnarray}
    (\delta x) (\delta {p_x}^H) \ge \frac{\hbar}{2} \left( 1 + \left( \frac{3}{2 m^2 c^2} - \frac{(\Delta L)^2}{2\hbar^2} \right) \Bigg( (\delta {p_x}^H)^2 + \langle \hat{p}_x \rangle^2 \Bigg) + \frac{3}{8 m^4 c^4} \langle {\hat{p}_x}^4 \rangle + \dots \right).
\end{eqnarray}
With the limit that $\Delta L \to 0$ and $\langle \hat{p}_x \rangle = 0$, the result of high-energy GUP is
\begin{eqnarray}
        (\delta x) (\delta {p_x}^H) &\ge& \frac{\hbar}{2} \left( 1 + \frac{3}{2 m^2 c^2} (\delta {p_x}^H)^2 + \frac{3}{8 m^4 c^4} \langle {\hat{p}_x}^4 \rangle + \dots \right), \nonumber \\ 
        &\ge& \frac{\hbar}{2} \left( 1 + \beta_m \frac{G}{\hbar c^3} (\delta {p_x}^H)^2 + \dots \right).
\end{eqnarray}
where $\beta_m = 3\hbar c/(2Gm^2)$ is the quadratic GUP parameter.
}

\end{document}